# Contribution of magnetism to the origin and stability of the rings of Saturn due to superconductivity of protoplanetary iced particles

Vladimir V. Tchernyi (Cherny) [1], Sergey V. Kapranov[2] and Andrey Yu. Pospelov[3]

[1] Modern Science Institute, SAIBR. Osennii blvd, 20-2-702. Moscow, 121614, Russia. Email: chernyv@bk.ru
[2] Moscow representative office of A.O. Kovalevsky Institute of Biology of the Southern Seas, Russian Academy of Sciences, Leninskiy Ave. 38/3, Moscow, 119991, Russia, Email: sergey.v.kapranov@yandex.ru
[3]Independent Researcher, Los Angeles, USA.

**Abstract:** It is demonstrated how Saturn rings may be originated due to interaction of iced particles (with II kind superconductivity) moving by chaotic orbits within protoplanetary cloud with magnetic field of Saturn after it appearance. Eventually all orbits of particles coming to the magnetic equator plane of Saturn where they locked within three-dimensional magnetic well due to minimum of the particles magnetic energy at the magnetic equator and quantum trapping by Abrikosov vortices. It also explains peculiarities of the rings pattern and many of observed phenomena.

## Introduction

After four missions to Saturn: Pioneer-11, Voyager-1 and -2, and outstanding Cassini (2204-2017) the answer of the rings origin problem made by Estrada, Durisen and Cuzzi (2017) is questionable: "After Cassini grand final, is there a final consensus on the ring origin and age?" Fridman & Gor'kavyi (1999) theory devoted to gravity defragmentation when asteroid was tidally disrupted passing nearby Saturn. Cassini found no iron in the rings. And Canup (2010) decided to use Titan-sized icy satellite instead of asteroid. But this theory could not be considered as entirely completed because it doesn't provide explanation of the following phenomena:

- Origin, evolution and dynamics of the rings
- Why rings are located within magnetic equator plane
– Considerable flattening and sharp edges of the rings system
– Why particles of the rings are separated and could collide
- Why there is a change of azimuth brightness of the Saturn *A* ring
– Why there are spokes in the Saturn *B* ring
– Why there is a roll-off of spectra at submillimeter wavelengths in Saturn *A*,*B*,*C* rings
– Why there is a wide band pulse radiation of the rings at 20 KHz – 40, 2 MHz range
– Why there is a radial dust flow of small particles (rain) falling down to planet due to gravity
– Why there is stability of particles within magnetic equator plane in the horizontal and vertical directions
– Why rings of planets appear only after asteroid belt
- Why Earth has no rings

Our model of the rings origin is based on Safronov (1972) theory of evolution of the protoplanetary cloud. Particles are rotating around Saturn in accordance with Kepler's law. We propose rings are result of the interaction of particles with appeared magnetic field of Saturn. The question is how all orbits of the particles may come to the magnetic equator plane.

### Magnetic model with II kind superconductivity of the iced particles

Low temperature of the environment near Saturn and presence of the planetary magnetic field bring us to idea of the importance of particles diamagnetism and superconductivity. Cassini found rings particles consist of 93% of ice and 7% of carbon. Babushkin et al (1986), Yen & Gao (2015) demonstrated diamagnetism of ice and Cote et al (1998) demonstrated superconductivity of carbon.

In 2011 Deutscher, Almog et al demonstrated how small size of superconductor conjugated together with exceptional large scale diamagnetic material is trapped in a surrounding magnetic field and it can be made to hover over a magnet in any position with any movements, and entire sample behave like a type II superconductor. In such a case the Meissner effect is observed even for sufficiently weak magnetic field.

An interesting idea of the rings origin is coming from the analogy with experiment of general electromagnetism physics. There is a similarity of the picture of Saturn's rings to the picture of iron particles created dense and rarefied regions in a nonuniform magnetic field near by the magnet bar on the table (Fig. 1). Strips of the iron particles density around magnet bar it is pure magnetic phenomenon.
Now will demonstrate how disk of the rings particles was originated from the particles of the protoplanetary cloud due to diamagnetism of ice and superconductivity of carbon which both together created particles of the II kind superconductor (Tchernyi & Kapranov, 2019).

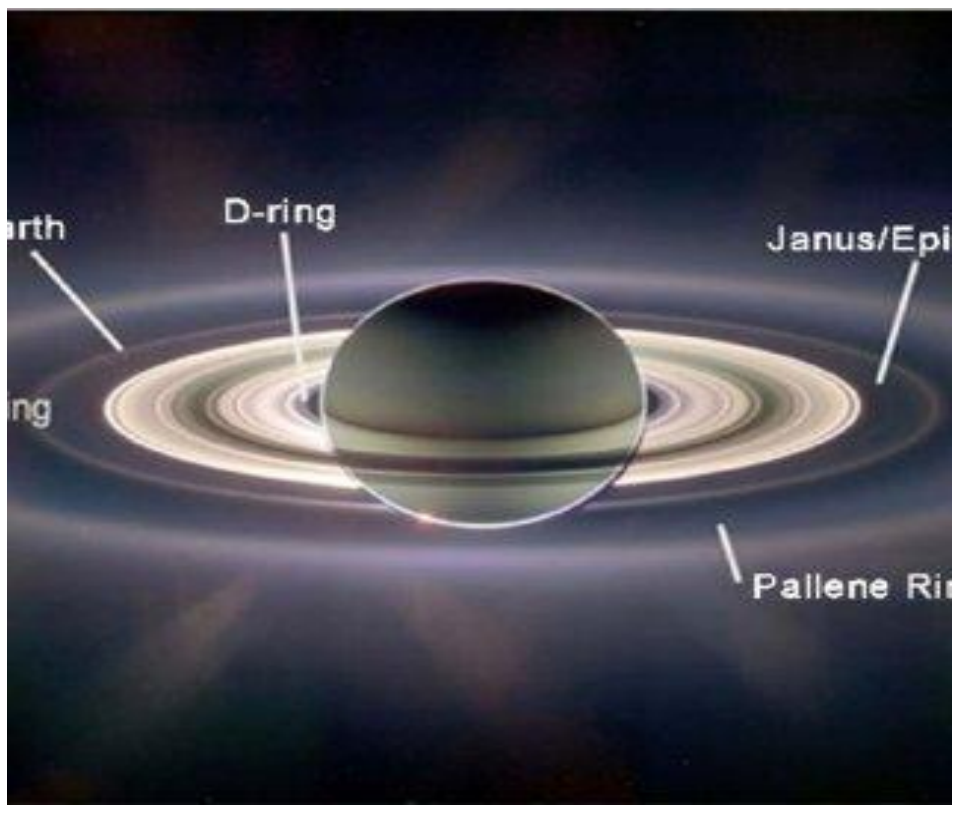

(a)

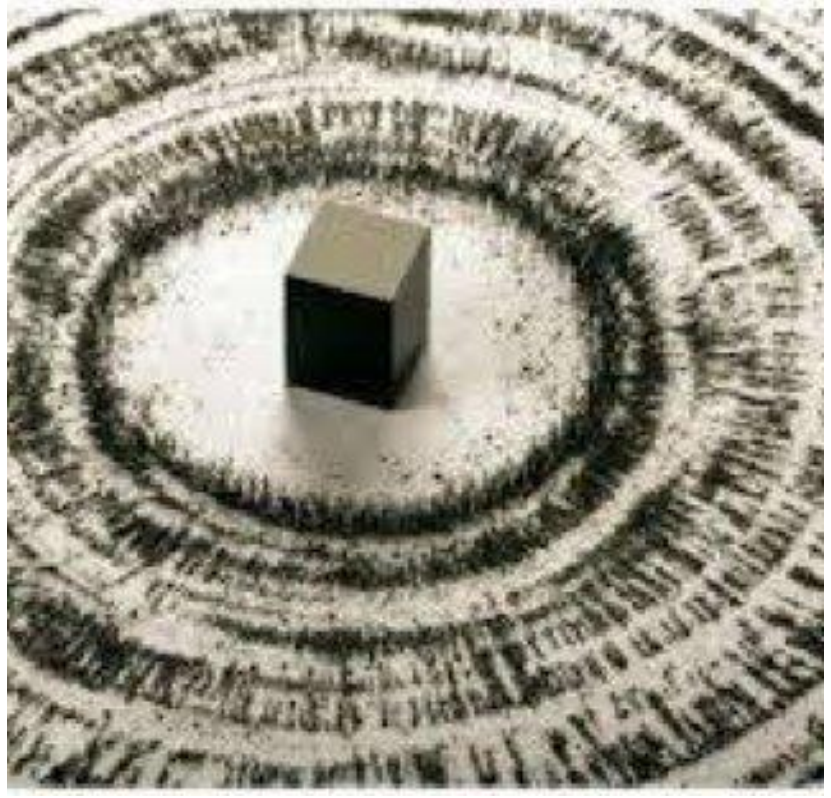

(b)

**Fig. 1.a.** https://saturn.jpl.nasa.gov/resources/. Image and Title Credit: NASA/JPL/JHUAPL/University of Colorado/Central Arizona College/SSL. Saturn's rings system.
**Fig.1.b.** https://wadevenden.files.wordpress.com/2014/07/magnet.jpg?w=500. Iron particles create dense and rarefied regions in nonuniform magnetic field around magnet bar on the table.

**Origin of the rings from superconducting particles of protoplanetary cloud due to Meissner phenomenon**

After origin of Saturn's magnetic field all iced particles of the protoplanetary cloud begin to interact with it due to appearance of additional force of diamagnetic expulsion of particles and all of them become involved in an additional azimuth-orbital motion. If coordinate system originates from the center of planet and z axis is directed along magnetic moment μ and orthogonal to equator, the particle's magnetic energy is (Tchernyi & Chensky, 2005)

$$U_H = \frac{R^3 \mu^2}{r^6}(3\cos^2\theta + 1)$$

Superconducting particle of radius *R* located within the protoplanetary cloud at the distance *r*, the angle *θ* between the vector *r* and the axis z. The particle's magnetic energy has a minimum value when *r* is in the magnetic equator plane, cos θ = 0. In case of one particle its orbit can be disturbed only by magnetic field. In case of big amount of particles after transient time estimated as 1000 years or more, collisions between particles will compensate their azimuth-orbital movements, and as a result all orbits of the particles of the protoplanetary cloud should come together to the magnetic equator plane and create highly flattening disc around the planet. Within the disc of rings all particles will become located on the Kepler's orbit where there is a balance of gravity, centrifugal force and diamagnetic force of expulsion.

**Repelling and collision of superconducting particles inside the rings**

For particles with magnetic moments $\mu_{1z}$ and $\mu_{2z}$ located on the same plane, z=0, the interaction energy is

$$U = \frac{\mu_{1z}\mu_{2z}}{\rho^3}$$

From it follows that both particles will repel each other and they will maintain a separate distance between them. If particles are located on the same axis but on different planes, the expression for the interaction energy is

$$U = -\frac{\mu_{1z}\mu_{2z}}{|z|^3}$$

Here particles could attract each other, collide or stick together and form bigger clusters or lumps of ice.

## Electromagnetic phenomena related to superconductivity of the Saturn's rings particles

### Thin width and sharp edges of the rings

Superconducting particles of the protoplanetary cloud collapsed into the stable rings system. The force of the diamagnetic expulsion forms sharp edges of the rings. The sudden breaks in the rings will be stabilized by this force too. Magnetic field of the rings disc is nonuniform because of each particle is push out own magnetic field. Magnetic field lines try to avoid particles. Gradient of density flow of the magnetic field repel particles of each other, and it also cleans the gaps, and forms a rigid thin structure of separated rings. The density of the magnetic flow inside each ring is lower than in around space. The difference of density of the magnetic flow is cause the magnetic pressure which is directed inward of each ring. Therefore the rings have sharp edges (Tchernyi & Pospelov, 2007).

### Radial dust flow from the rings

Submicron particles which lost character size of superconductors will fall down to the Saturn due to gravity. Also this process is possible due to particles collisions and fluctuation of the magnetic field.

### Wide band pulse radiation of the rings

Pulse radiation at the 20 KHz - 40, 2 MHz could take place if distance between particles is ~10-8 m or there is a point contact. Superconducting transition can occur when electrons is tunneling through it. This type of weak link can generate electromagnetic radiation like in case of non-stationary Josephson phenomenon for superconductors.

### The azimuth brightness of the *A* ring

If superconducting particle is in the magnetic field, a magnetic moment directed opposite to the external field is induced. Magnetization is not along the external magnetic field but in the opposite direction. Superconducting particle in the form of the rod tries to place itself perpendicularly to the magnetic field lines. Due to Maeno's theory of ice (1981) growing snowflakes below - $22^0$ C take the form of prisms. And the prism of the superconducting ice particle will be oriented perpendicularly to the field lines of the poloidal and toroidal components of the magnetic fields of Saturn. Now it is clear the variable azimuth brightness of the Saturn's *A* ring may be related to orientation of the elongated ellipsoid of the superconducting particles versus the direction of the planetary magnetic field.

**Spokes within *B* rings system**

Spokes in the Rings *B*

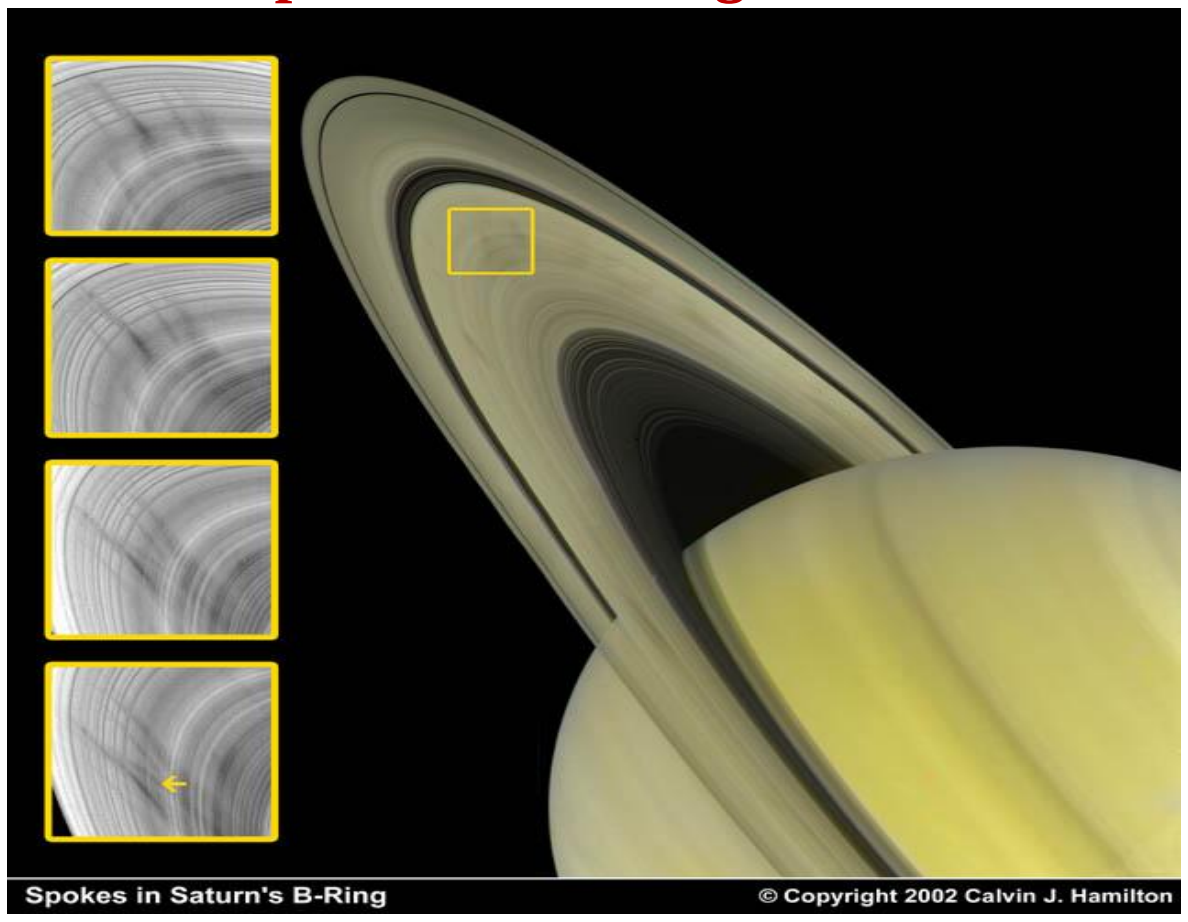

**The size of spokes ~ 104km x103 km. Periodicity ~ 640 min is as Period of Magnetic Field Rotation. If Particles Enter into AnomalyArea, the Change of Diamagnetic Expulsion Force is Changed Their orbits. Effect is Noticeable Due to Dense Fog of Particles. Ring B is bright but has less Transparency than Swimming pool: M. Hedman, P. Nicholson, Icarus, 22.01.2016**

Fig. 2. **Spokes within *B* rings system**

Hedman & Nicholson (2016) found Saturn B ring has three times less matter than previously thought. Nicholson: "The best analogy is something like a fog over the meadow may seem less transparent and empty, than a water-filled pool, which has a much higher density than the fog". The spokes of the B ring are located almost radially. Their size is ~ 104 km along the radius and about 103 km along the orbit. The spectral radiation power has periodicity ~ 640.6±3.5 min which almost coincides with the period of rotation of the magnetic field of Saturn (639.4 min). There is correlation of maxima and minima of activity of spokes with spectral magnetic longitudes which is connected to presence or absence of the radiation of Saturn's Kilometric Radiation. When superconducting particles enter into anomalous region the diamagnetic expulsion force that is applied to the particles changes its value. The particles then begin to change their orbit. For the huge number of particles, for the external observer, this process appears as the turbulent cloud stretched along the radius in the form of spokes. After passing anomaly, superconducting particles return to their prior orbit and appearance of the rings is recovered.

**Frequency anomalies of thermal radiation of the rings**

Cassini Composite Infrared Spectrometer (CIRS) spatially resolved Saturn's main rings in the far-infrared, measuring the spectrum between 25 μm and 0.5 mm (Spilker et al, 2005). A spectral roll-off below 200 μm for each of the A, B and C rings was found (Fig.3.a) The data temperatures and emissivities for each ring were derived. Interpretation of Cassini CIRS spectral roll-off in Saturn's rings has encountered difficulties. We found explanation (Pospelov, Tchernyi, 2019). From the comparison of Fig.3.a and Fig.3.b we see the same behavior of the spectral dependence of Saturn's rings and superconductor. It follows the Saturn's rings particles could possess superconductivity.

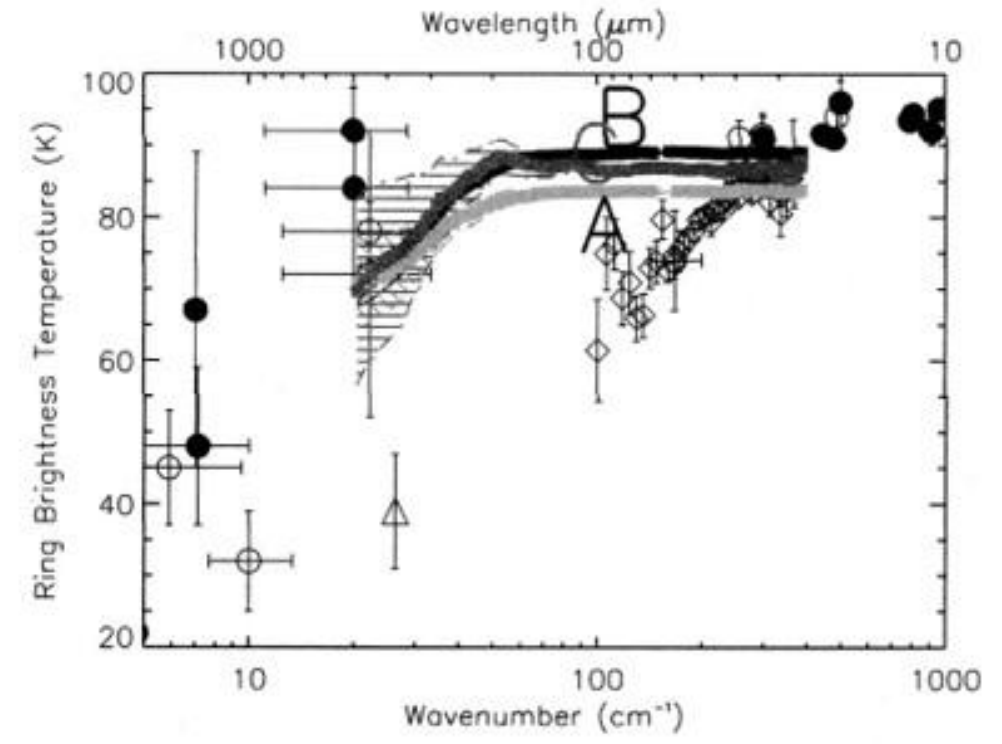

(a)

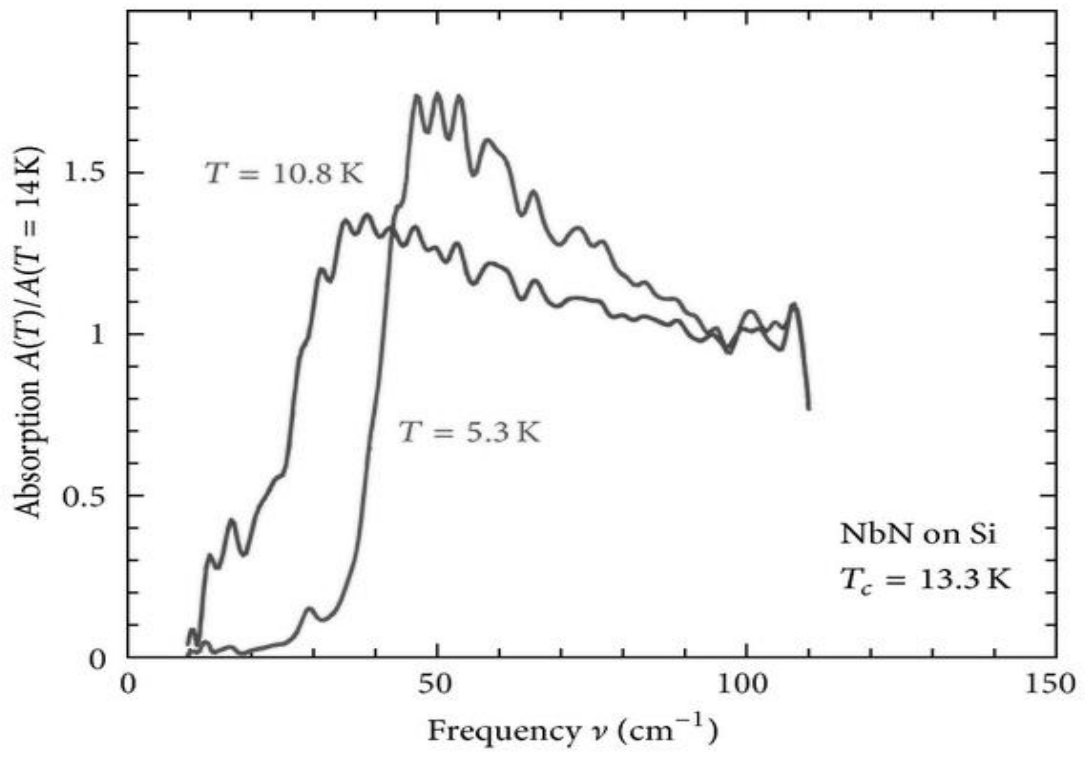

(b)

**Fig. 3.a.** Brightness temperatures of the A, B and C rings as a function of spectra are shown by Spilker et al (2205). **Fig. 3.b.** Absorption spectrum of superconductor. M. Dressel. Electrodynamics of metallic superconductor. V. 2013.

**Quantum phenomenon of stability of the Saturn rings particles within magnetic equator plane**

Deutscher, Almog et al (Chicago ASTC 2011) demonstrated that even small amount of superconducting matter may hold more than 70,000 times its own weight. It is based on theoretical discovery of Abrikosov (1957) of the vortex structure as a quantum nature of the II kind of superconductor. In case of Saturn's rings we do have similar situation (Tchernyi, Pospelov, 2018). The rings particles contain 93% of ice and only 7% of carbon. It means that II kind of superconductivity is responsible for the rings particles disk formation due to contain only 7% of carbon.

An interaction of II kind superconductor with external magnetic field is a quantum physics phenomenon. The magnetic field inside superconductor behaves like a quantum particle, like a quantum object. And as a fact magnetic field is magnetized inside superconductor and strands of lines of the magnetic field remain inside superconductor, they get trapped inside superconductor (Fig.4). Some of the flux line becomes to be pin, and they never could take move. The reason for that is that superconductor doesn't like magnetic fields lines moving around. What it actually does, it locks them in place. By doing that it blocks itself. The action of locking prevents superconducting particles of the rings from moving within the disc. It is quantum trapping, quantum locking and quantum levitation. Also the action of trapping prevents the disc itself from moving in horizontal directions. For the Saturn disk of rings the problem of stability of the pinning structure of the superconducting iced particles is important. Stability of mathematical solution of the Ginzburg-Landau equation for superconductor was presented by Sigal & Tzaneteas (2013, 2016).

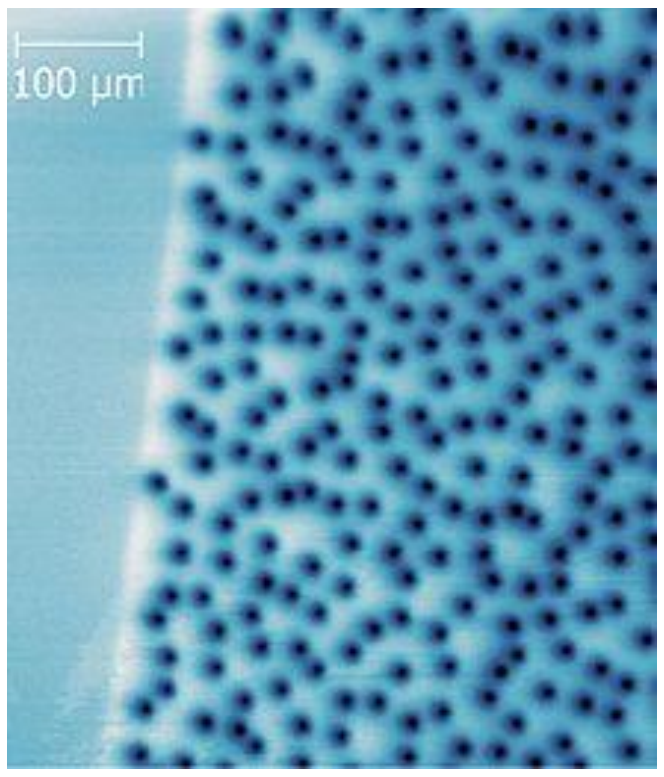

(a)

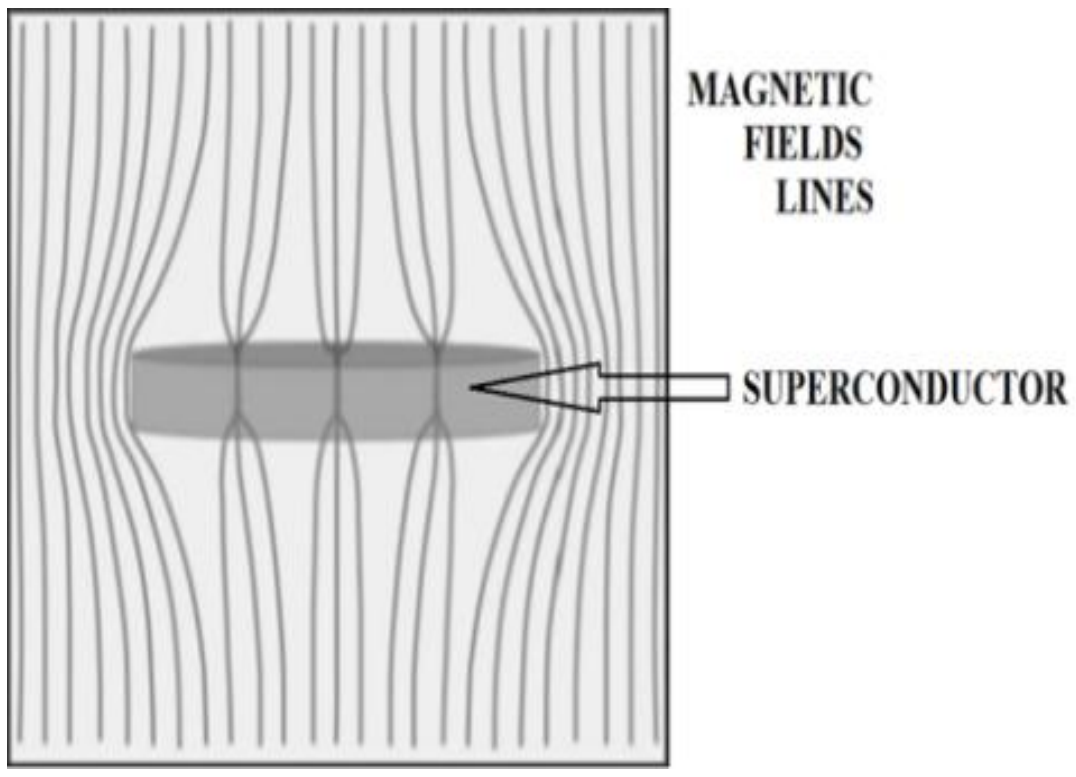

(b)

**Fig. 4.a.** Photo of the surface of a 200-nm-thick YBCO superconducting film with flux pinning structure. Huge amount of magnetic field lines penetrating sample of superconductor (vortices) and hold it against any horizontal movements, Wells et al (2015).
**Fig. 4.b.** Superconducting iced particle of the rings which is located along the horizontal equator plane is interacting with vertical Saturn magnetic field lines. Some of magnetic lines will be expelled outside of the sample of II kind superconductor. And some others will remain inside superconducting particle and it locks the magnetic field flux lines inside itself.

Consequently superconducting iced particles will be locked within the rings disc at the magnetic equator plane by the Saturn magnetic field due to phenomena of quantum locking, quantum trapping and quantum levitation. Saturn axis is tilted with respect to the plane of its orbit around the Sun, and Saturn rings are located in the magnetic equator plane. It is also another conformation of important role of Saturn's magnetic field for the rings origin.

The rings disk itself will be suppressed by the magnetic pressure from both sides along the z axis, because along a meridian the magnetic energy becomes to be bigger on the distance from the minimum value at the magnetic equator. Eventually we have a magnetic well of the rings disc which is disturbing picture of Saturn magnetic field lines in the area of magnetic equator, Fig.5.

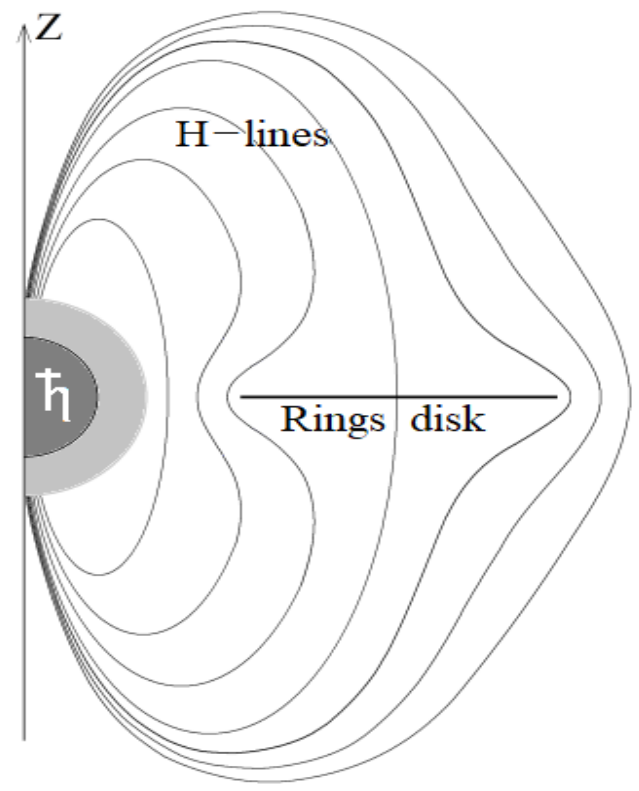

**Fig. 5.** Rings disk of superconducting particles is disturbing magnetic fields lines of Saturn.

**Conclusion**

It is demonstrated that presence of diamagnetism of ice and superconductivity of carbon from which consist of iced particles of Saturn's rings is allowed to consider each particle as II kind superconductor. Then Saturn's rings can be formed due to interaction of iced particles moving inside the protoplanetary cloud by chaotic orbits with magnetic field of Saturn after its origin and appearance of additional force of diamagnetic expulsion of particles. Eventually after collisions of particles all their orbits coming to the magnetic equator plane of Saturn where they locked within three-dimensional magnetic well due to minimum of particles magnetic energy at the magnetic equator and quantum trapping of particles by Abrikosov vortices.
Finally we've received explanation of the origin of Saturn's rings from superconducting particles of protoplanetary cloud since appearance of magnetic field of Saturn. To estimate age of the rings it is necessary take into account time after origin of magnetic field needed for all particles orbits moved to magnetic equator plane.

Presented theory doesn't destroy any contribution of gravity, dusty plasma and magnetohydrodynamic phenomena. This model explains peculiarities of the pattern of Saturn's rings system and many of observed physical phenomena.

It may be useful for understanding of the rings of planets behind the asteroid belt and for exoplanets.
This theory also attracts attention to the problem of studying natural superconductivity in space.

At the same time we have to note this theory is underlined of important role of magnetic field of the Saturn's rings origin. There is similarity of physical situation of the rings origin and formation of the iron particles stripes and gaps around bar of magnet on laboratory table where gravity of the Earth doesn't play the main role because if there is no magnet bar then there is no pattern with stripes.

We can present transformation of protoplanetary cloud filled with superconducting iced particles into disk of rings consisting of superconducting particles after appearance of the magnetic field of Saturn due to interaction of iced particles with it within cloud as it is on Fig.6.

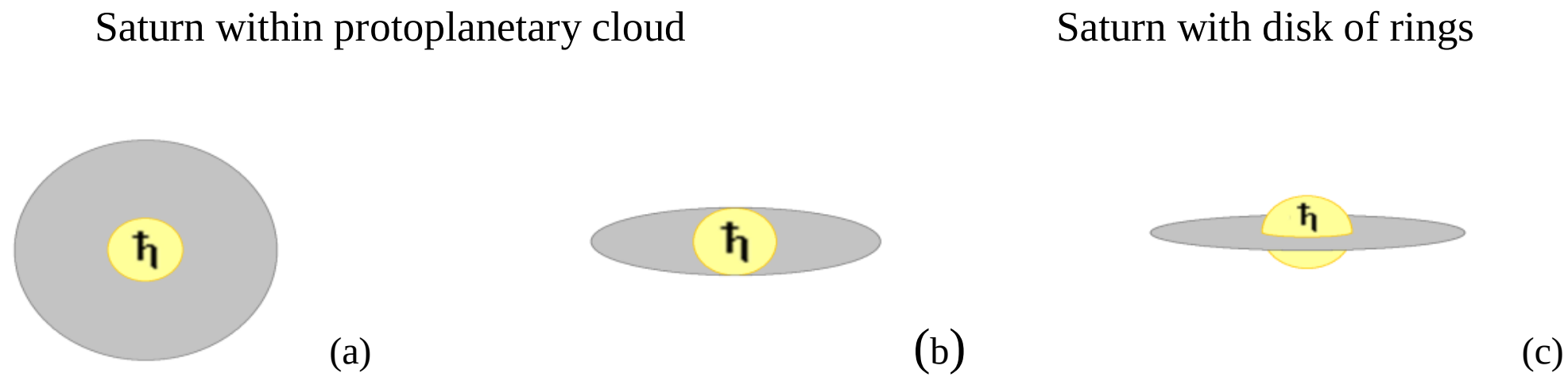

**Fig. 6.** Transformation of Saturn's protoplanetary cloud filled with superconducting iced particles into disk of rings consisting of superconducting particles after appearance of Saturn's magnetic field and interaction of it the iced particles: from (a) >> (b) >> (c).

This work was presented at the meeting of aas235 Honolulu, Jan. 4-8, 2020.
Part of work contributed by S.V. Kapranov was supported by the state assignment AAAA-18-118021350003-6 of A.O. Kovalevsky Institute of Biology of the Southern Seas, Russian Academy of Sciences.

There is 4-min. movie related to presented theory made by someone on July 31, 2019 after our paper arXiv.org/abs/1907.07114 on July 8, 2019: The mystery of Saturn Rings Solved By Magnetism?
https://www.youtube.com/watch?v=AI6AaMJoR4A